\documentclass[12pt, centerh1]{article}

\usepackage{amsmath,natbib}
\usepackage{graphicx,subfigure}
\usepackage{verbatim}
\newcommand{\secref}[1]{Section~\ref{#1}}
\newcommand{\appref}[1]{Appendix~\ref{#1}}
\newcommand{\tabref}[1]{Table~\ref{#1}}
\newcommand{\figref}[1]{Figure~\ref{#1}}

\newcommand{\pa}{\pAB }
\newcommand{\pAB}{P(A,B)}
\newcommand{\pc}{\pAnotB }
\newcommand{\pb}{\pnotAB }
\newcommand{\pd}{\pnotAnotB }
\newcommand{\pA}{P(A)}
\newcommand{\pnotA}{P(\overline A)}
\newcommand{\pB}{P(B)}
\newcommand{\pnotB}{P(\overline B)}
\newcommand{\pnotAB}{P(\overline A, B)}
\newcommand{\pAnotB}{P(A,\overline B)}
\newcommand{\pnotAnotB}{P(\overline A,\overline B)}

\newcommand{\gini}{{\mathcal{G}}}
\newcommand{\q}{{\mathcal{Q}}}
\newcommand{\lift}{{\mathcal{L}}}
\renewcommand{\cos}{{\mathcal{C}}}

\newcommand{\sgini}{\gini^*}
\newcommand{\sq}{\q^*}
\newcommand{\scos}{\cos^*}
\newcommand{\slift}{\lift^*}
\newcommand{\Rule}{A\Rightarrow B}

\newcommand{\pack}[1]{{\tt{#1}}}

\newcommand{\R}{\pack{R}}

\begin{document}


%

\title{Standardizing Interestingness Measures\\ for Association Rules}
\author{Mateen Shaikh, Paul D. McNicholas\thanks{Department of Mathematics \& Statistics, University of Guelph, Guelph, Ontario, N1G 2W1, Canada. E-mail: pmcnicho@uoguelph.ca.}, M. Luiza Antonie and T.\ Brendan Murphy}
\date{Department of Mathematics \& Statistics, University of Guelph}
\maketitle

\begin{abstract}

Interestingness measures provide information that can be used to prune or select association rules. A given value of an interestingness measure is often interpreted relative to the overall range of the values that the interestingness measure can take. However, properties of individual association rules restrict the values an interestingness measure can achieve. An interesting measure can be standardized to take this into account, but this has only been done for one interestingness measure to date, i.e., the lift. Standardization provides greater insight than the raw value and may even alter researchers' perception of the data. We derive standardized analogues of three interestingness measures and use real and simulated data to compare them to their raw versions, each other, and the standardized lift.


\end{abstract}

%




\section{Introduction}
\label{intro}

Association rules have been the subject of considerable research since the seminal paper by \cite{agrawal93}. They are used to identify trends and patterns in transactional data such as market-basket analyses, where researchers investigate common purchasing trends. Once mined, association rules are often compared to each other using some interestingness measure. Different interestingness measures quantify different characteristics of an association rule \citep[see][for discussions]{piatetsky91,freitas99,tan02,geng06}. 
Each interestingness measure takes values on some range; however, to only consider the value of an interestingness measure without reference to this range can be misleading, cf.\ \secref{sec:lift}. To overcome this issue, interestingness measures can be standardized to provide values that reflect the value achieved relative to the values that are obtainable. To date, only one interestingness measure has been standardized --- the lift \citep{mcnicholas08c} --- and so the effects of standardizing interestingness measures are not well understood in general. Herein, we standardize three additional interestingness measures and explore the corresponding effects through the analysis of three data sets.

Association rules are defined in the context of a non-empty set of items~$I$, called the itemset. An association rule is a statement of the form $A \Rightarrow B$, where $A \subset I, B \subset I$ such that $A\neq\emptyset$, $B\neq\emptyset$, and $A\cap B=\emptyset$. The set $A$ is called the antecedent and $B$ is called the consequent. Association rules are mined over many transactions. A transaction is a single instance wherein each item is present or absent.
%
The support of $X\subset I$, denoted $P(X)$, is the proportion of transactions that contain every item in $X$. Consider a rule $\Rule$. $\pA$ denotes the support of the antecedent, $\pB$ denotes the support of the consequent, and $\pAB$ (or equivalently $P(\Rule))$  denotes the support of the rule itself. The confidence of a rule is 
\begin{equation}\label{conditionalprobability} P(B\mid A)=\frac{\pAB}{\pA}. \end{equation} 

There are many instances when the absence of items within transactions is also of interest. $P(\overline X)$ denotes the proportion of transactions that do not contain every item in $X\subset I$. For example, one may be interested in the proportion of transactions that contain every item in $A$ but do not contain the items in $B$, i.e., the rules $A\Rightarrow\overline{B}$ and $\overline{B}\Rightarrow A$. 
The set $\overline{A}$ is called the negation of $A$. Although we do not explicitly consider rules containing negations herein, our calculations are more concise and understandable using this notation.

\section{Interestingness Measures} \label{sec:interestingness measures}

\subsection{General Remarks and Properties}
There are many different interestingness measures for association rules and some are more suitable for some purposes than others. \cite{piatetsky91} outlined three important properties every interestingness measure should have. The first property (Prop.~1) is that the interestingness measure should achieve the value $0$ at independence, i.e., when $\pAB=\pA\pB$. A non-zero constant, if known in advance, satisfies the same purpose, e.g., a value of one indicates independence for lift (cf.\ Section~\ref{sec:lift}). The second desirable property (Prop.~2) is that the value of the interestingness measure should be larger when $\pAB$ is larger but $\pA$ and $\pB$ remain the same. The third property (Prop.~3) is related to the second: the interestingness measure should be smaller when either $\pA$ or $\pB$ increases but $\pAB$ is unchanged, or when $\pB$ increases but $\pA$ and $\pAB$ are unchanged. 

In addition to these three properties, five others have been identified by \cite{tan02} and considered by \cite{geng06}. These properties are not necessarily desirable but can be important to help distinguish interestingness measures from one another. One property is symmetry (Prop.~4): the interestingness measure remains the same when the antecedent and consequent are interchanged. Another property is antisymmetry under one negation (Prop.~5), i.e., the measure's value is multiplied by $(-1)$ when the antecedent is replaced by its negation or the consequent by its negation. A related property (Prop.~6) is symmetry under two negations, i.e., the measure's value is multiplied by $(-1)$ when the antecedent is replaced by its negation and the consequent by its negation. Prop.~5 trivially implies Prop.~6; however, some interestingness measures, such as the odds ratio and the Gini index, satisfy Prop.~6 but not Prop.~5. Another suggested property (Prop.~7) is that the interestingness measure should achieve the same value under row-wise or column-wise scaling (or both). The final property is null invariance (Prop.~8): additional transactions where neither the antecedent nor the consequent appear do not change the value of the interestingness measure.

A total of 21 interestingness measures are considered by \citet{tan02}. These are discussed in the specific context of association rules by \citet{geng06}, who outline methods of choosing suitable interestingness measures. Part of their analysis includes identifying interestingness measures that satisfied the three properties of \citet{piatetsky91} as well as the five additional properties discussed by \citet{tan02}. This analysis includes a table (their Table 6) that shows whether or not each of the 21 interestingness measures satisfies these eight properties. In an attempt to group these interestingness measures, their analysis also contains a figure (their Figure~3) identifying levels of similarity, by correlation, between interestingness measures. It is from this figure, where all values of support are considered, that we identify at least three separate groupings of interestingness measures. Interestingness measures that appear in the same group often do not satisfy the same subset of the eight properties but are well correlated, in that they have correlation values of least 0.85 with any other interestingness measure in the same group. Because of this similarity, we choose one representative from each of the three groups to analyze: cosine similarity, Yule's Q, and the Gini index. \tabref{tab:properties} shows a summary of the three interestingness measures in the same fashion as Table~6 from \citet{tan02}.
\begin{table}[!h]
\caption{\label{tab:properties} Properties of four selected interestingness measures.}
\addtolength{\tabcolsep}{-3pt}
\begin{tabular}{l|cccccccc}
\hline
	& Prop.~1& Prop.~2& Prop.~3& Prop.~4& Prop.~5& Prop.~6& Prop.~7 & Prop.~8\\
\hline
Lift			& Y	& Y 	& Y 	&  Y	& N 	&  N	& N 	& N	\\
Cosine Similarity	& N	& Y 	& Y 	&  Y	&  N	&  N	& N 	& Y	\\
Yule's Q		& Y	& Y 	& Y 	&  Y	&  Y	&  Y	& Y 	& N	\\
Gini Index		& Y	& N 	& N 	&  N	&  N	&  Y	& N 	& N	\\
\hline
\end{tabular}
\end{table}

\subsection{Lift} \label{sec:lift}
The lift of a rule, which has also been called the `interest' of a rule \citep{brin97}, is denoted here as $\lift(\Rule)$ and is defined as
\begin{equation}\label{eq:lift}
	\lift(\Rule)=\frac{\pAB}{\pA\pB}.
\end{equation}
Lift takes on non-negative values and equals one when the antecedent and consequent are statistically independent. As an example of when a measure such as lift can be problematic, we consider an instance where the support of the antecedent of a rule, $\Rule$, is 0.5 and the overall lift value for this rule is 1.95. Because $\pA=0.5$, we know that the value of lift for this rule cannot exceed 2. Suppose, also, that for another rule, $C\Rightarrow D$, the lift of the rule is 1.95 but the support of the antecedent, $P(C)$, is $0.1$. In this second rule, the lift value was not bounded above by 2 but  by 10. Even though these two rules have the same lift value of 1.95, the interpretation of the two lift values should not be the same because the maximum attainable value is different. Such discrepancies occur when using raw interestingness measures; standardizing lift \citep{mcnicholas08c} addresses the issue described here. In this example, standardized lift values for rules $\Rule$ and $C\Rightarrow D$ are $0.975$ and $0.195$, respectively, all else being equal. 

\begin{figure}[!h]
	\vspace{-1.7in}
	\centering
	\includegraphics[width=0.8\textwidth]{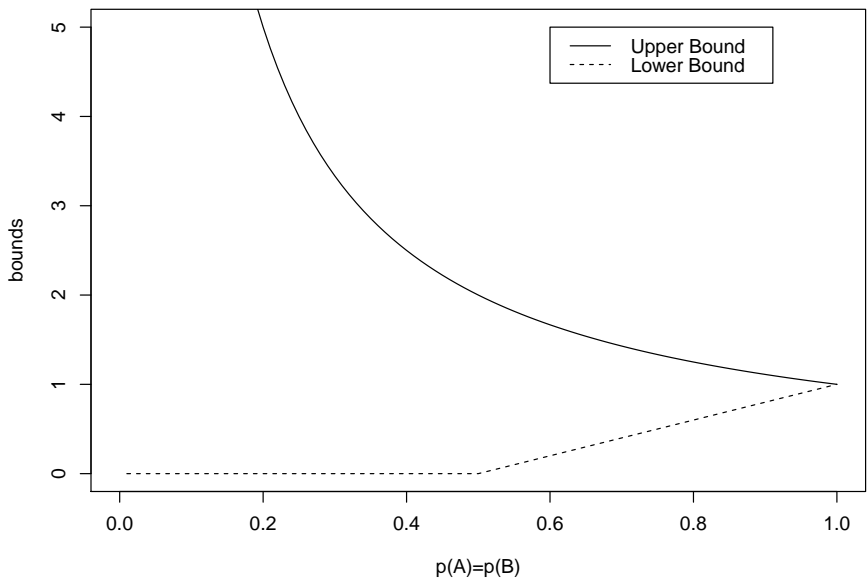}
	\vspace{-1.9in}
	\caption{\label{fig:liftbounds}The bounds on lift when $\pA=\pB$.}
\end{figure}

To show just how bounded lift is, consider the values of lift where $\pA=\pB$. We can see (\figref{fig:liftbounds}) that for larger supports, the value of lift is bounded near one. Without any other considerations, the small differences in the values of lift between rules may be overlooked; however, these differences may be more significant after considering the narrow range of possible values. Though we only highlight this issue in the case when $\pA=\pB$, such problems exist in many cases.

The standardized lift of a rule, $\slift(\Rule)$, is obtained by finding the suitable upper bound, $\upsilon$, and lower bound, $\lambda$, in the following definition:
\begin{equation*}\label{eq:standardizedlift}
	\slift(\Rule)=\frac{\lift(\Rule)-\lambda}{\upsilon-\lambda}.
\end{equation*}
\citet{mcnicholas08c} derive the following values of $\upsilon$ and $\lambda$: 
\begin{equation*}\begin{split}
	\upsilon &= \frac{1}{\max\{\pA,\pB\}},\\
	\lambda &= \max \left\{ \frac{\pA+\pB-1}{\pA\pB}, \frac{4\sigma}{(1+\sigma)^{2}}, \frac{\sigma}{\pA\pB}, \frac{\kappa}{\pB} \right\}.
\end{split}\end{equation*}
These values are functions of the minimum support and confidence thresholds, denoted in this paper by $\sigma$ and $\kappa$, respectively. These thresholds are assumed to be at least ${1}/{n}$, where $n$ is the number of transactions. 
Analogous approaches are used herein to standardize three other interestingness measures.


\subsection{Cosine Similarity} \label{sec:cosine}
	The definition of the cosine similarity resembles the definition of the lift of a rule and is:
		\begin{equation}
				\cos(\Rule)=\frac{\pAB}{\sqrt{\pA\pB}}. \label{eqn:cos}
		\end{equation}
The cosine similarity lies on the closed unit interval and no constant value is achieved when the antecedent and consequent are statistically independent of one another. In such cases, the cosine similarity attains the value $\sqrt{\pA\pB}$. The standardized cosine similarity of a rule $\Rule$ is given by
$$
	\scos(\Rule)=\frac{\cos(\Rule)-\lambda}{\upsilon-\lambda},
$$
where 
\begin{equation*}
\upsilon=\min\left\{\sqrt{\frac{\pA}{\pB}} , \sqrt{\frac{\pB}{\pA}}\right\}
\end{equation*}
and 
\begin{equation*}
\lambda=\max\left\{
\frac{2\sigma}{1+\sigma},
\frac{\sigma}{\sqrt{\pA\pB}},
\frac{\pA+\pB-1}{\sqrt{\pA\pB}},
\sqrt{\kappa\frac{\sigma}{\pB}},
\kappa\sqrt{\frac{\pA}{\pB}}
\right\}.
\end{equation*}
 Derivations for $\upsilon$ and $\lambda$ are found in \appref{sec:cosderivation}.

\subsection{Yule's Q}\label {sec:yule}

	Some interestingness measures take supports of negations into account. One such interestingness measure is Yule's Q \citep{yule03}, which is based on the odds ratio. Unlike the odds ratio, however, Yule's Q remains defined in all cases where \mbox{$\pA>0$} and $\pB>0$. Yule's Q takes values on $[-1,1]$ and is defined as 
		\begin{equation}
				\q(\Rule)=\frac{\pa\pd-\pb\pc}{\pa\pd+\pb\pc}. \label{eqn:Q}
		\end{equation}
This interestingness measure obtains a value of zero when the antecedent and consequent are statistically independent. Positive values indicate that the antecedent and consequent are more likely to occur together than if they were independent, and negative values indicate that the antecedent and consequent are less likely to occur together than if they were independent. Akin to lift and the cosine similarity, Yule's Q is symmetric. 

The standardized Yule's Q of a rule $\Rule$ is
$$\sq(\Rule)=\frac{\q(\Rule)-\lambda}{\upsilon-\lambda},$$
where $\upsilon = 1$ and 
\begin{equation*}\begin{split}
\lambda = 	\max&\left\{  -1, \frac{\sigma-\pA\pB}{\sigma+\pA\pB-2\sigma(\pA+\pB-\sigma)} , \right.\\&
\qquad\qquad\qquad\qquad\qquad\left.\frac{\kappa-\pB}{\kappa+\pB-2\kappa(\pA+\pB-\kappa\pA)}  \right\}.
\end{split}\end{equation*}
Derivations for $\upsilon$ and $\lambda$ are found in \appref{sec:yulederivation}.

\subsection{Gini Index}\label{sec:gini}
The Gini index is the only measure discussed here that is not symmetric. That is, the Gini index for the rule $\Rule$ may not achieve the same value as the Gini index for the rule $B\Rightarrow A$.
Used in a variety of contexts, the Gini index quantifies impurity and is defined as
\begin{equation}
		\begin{split}\label{eq:gini}
				\gini(\Rule)&=\pA\left[P(B\mid A)^2+P(\overline B \mid A)^2\right]\\&\qquad+\pnotA[P(B\mid \overline A)^2+P(\overline B \mid \overline A)^2]-\pB^2-\pnotB^2.
		\end{split}
\end{equation}
	This interestingness measure achieves a value of zero when the antecedent and consequent are statistically independent, and achieves a value between zero and 0.5 when the antecedent and consequent are not independent.
The standardized Gini index of a rule $\Rule$ is
$$
	\sgini(\Rule)=\frac{\gini(\Rule)-\lambda}{\upsilon-\lambda},
$$
where
\begin{equation*}\upsilon=\begin{cases}
	\hfill2\frac{(\min\{\pA,\pB\}-\pA\pB)^2}{\pA(1-\pA)} & \text{if } \pAB\geq \pA\pB, \\[1em]
	2\frac{(\max\{\sigma,\kappa\pA,\pA+\pB-1\}-\pA\pB)^2}{\pA(1-\pA)} & \text{if } \pAB< \pA\pB,
\end{cases}\end{equation*}
and 
\begin{equation*}\lambda=\begin{cases}
	2\frac{(\max\{\sigma,\kappa\pA,\pA+\pB-1,\pA\pB\}-\pA\pB)^2}{\pA(1-\pA)} & \text{if } \pAB\geq \pA\pB,\\[1em]
	\quad 0 & \text{if } \pAB< \pA\pB.
\end{cases}\end{equation*}
Derivations  for $\upsilon$ and $\lambda$ are given in \appref{sec:giniderivation}. 

\section{Experimental Results and Evaluation}

We explore the effects of standardizing interestingness measures with three different data sets. In the first data set, values are obtained for lift, cosine similarity, Yule's Q, and the Gini index. We see that values of each interestingness measure are clearly reordered when standardized, making previously unremarkable rules more noteworthy. The second data set that we analyze is comparably more sparse, and comparing the effect of standardizing interestingness measures with results from the first data set is interesting. We also consider a third data set that is comprised of completely simulated, random data, with independent items. These data draw a contrast between standardized interestingness measures and their non-standardized analogues. Specifically, they highlight that by taking into account the low probabilities of co-occurrences, standardized interestingness measures do not suggest that rules in random transactions are interesting.

For every data set, rules are mined using the Apriori algorithm \citep{agrawal94}. This method uses the Apriori principle, also known as downward closure; i.e., if a set of items is deemed to be frequent relative to a threshold, then every subset of this set will also be (at least as) frequent relative to the same threshold. Conversely, if a set of items is judged to be infrequent relative to a threshold, then every superset of this set will also be infrequent relative to the threshold. There are many alternatives to this algorithm, such as SETM \citep{houtsma95}; EClaT \citep{zaki00}, which divides the itemset into equivalence classes; frequency pattern trees \citep{han00}, which avoid many passes over the data; and variations to the Apriori algorithm itself such as AprioriTID, AprioriHybrid \citep{agrawal94}, and direct hashing and pruning \citep{park95}. Because the focus of our work is to explore the effects of standardizing interestingness measures, we employ the established and well-understood Apriori algorithm. Results for each data set can be visualized as a pair of scatterplot matrices (\appref{sec:scatterplotmatrices}); the first scatterplot matrix shows the raw interestingness measures and the second the standardized interestingness measures.

We note that the thresholds chosen in our exploratory analyses may not be typical for specific applications, which often use higher support thresholds and impose higher confidence thresholds. Such thresholds typically result in fewer rules, most often those indicating a positive correlation between the antecedent and consequent. To appropriately explore the effect of standardization, we intentionally keep these rules in our analysis when they existed, without the burden of a plethora of rules.

One method of comparing rule orderings is through Goodman and Kruskal's gamma \citep{goodman54}. In this context, it is calculated by determining the difference in the proportions of concordant pairs and discordant pairs while ignoring ties. Kendall's tau-b is a similar method of measuring the rankings that considers ties to be discordant, making it a more conservative description of relative ordering \citep[cf.][]{agresti02,blaikie03}. We use Kendall's tau-b herein. 
If the order of rules by raw interestingness measure is the same as the order of rules by the corresponding standardized interestingness measure, a value of $1$ results. If the order of one is the opposite of the other, a value of $-1$ results. A value of zero results when the two rule orderings have no apparent relationship, such as the case when one of the rule orders have been (uniformly) randomly ordered. In addition to calculating the tau-b value for all rules in each data set, we calculated tau-b by decile in each data set. Rather than numerically reporting these values, plots have been made available in Figure~\ref{fig:deciles} of Appendix~\ref{sec:deciles}. 

\section{Traffic Accidents}

Information from over 340,000 traffic accidents from Belgium's National Institute of Statistics \citep{geurts03} is analyzed by considering up to 572 factors that may have been involved. These factors include the type of road, intersection, features of the road, conditions, time, type of vehicle, direction, severity of accident, age of road user, obstacles, and other relevant criteria recorded by a police officer. As Table \ref{tab:accidentsrules} indicates, information identifying the actual factors has not been made available; rather, factors are simply enumerated. We set minimum support and confidence thresholds to 0.3 and 0.4, respectively. 
These thresholds were chosen because approximately 90,000 rules were produced when we considered entire rules that were no more than five items in length; we deemed this adequate for analysis.

\begin{figure}[!h]
	\centering\includegraphics[width=0.6\textwidth]{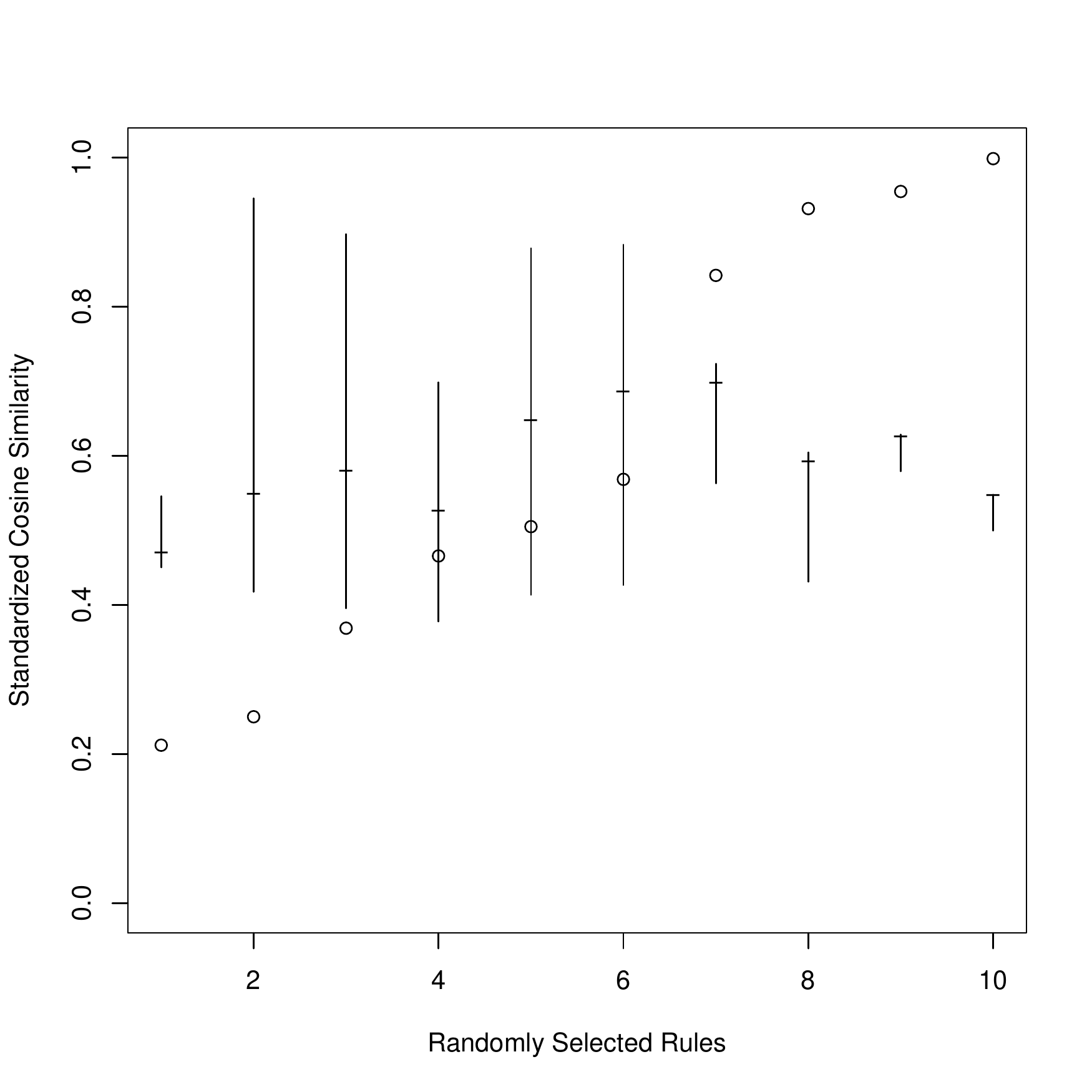}
	\vspace{-0.2in}
\caption{\label{fig:relative accidents} Cosine similarity values for randomly selected rules. Vertical lines represent the range of possible cosine similarity values for the rule, with the horizontal dash indicating the achieved value of the cosine similarity. The circle indicates the relative value of the standardized cosine similarity within the range of possible values.}
\end{figure}

To illustrate how just one interestingness measure is standardized, \figref{fig:relative accidents} shows ten randomly selected rules sorted by $\scos$. Each vertical bar represents the range of values of $\cos$ that were attainable by the rule based on the bounds discussed in \secref{sec:cosine}. The horizontal tick within the bar shows where the actual value of $\cos$ lies. The value of $\scos$ is the relative location of $\cos$ in the bar shown with circles. These values are overlain because the range of $\cos$ and $\scos$ are the same but the range of the raw and standardized interestingness measures are not coincident in general unless the range of the raw interestingness measure is also the unit interval. The rules are sorted by ascending values of $\scos$ and show that standardizing the interestingness measure is not simply a transformation of the raw interestingness measure.

\begin{figure}
	\centering\includegraphics[width=0.85\textwidth]{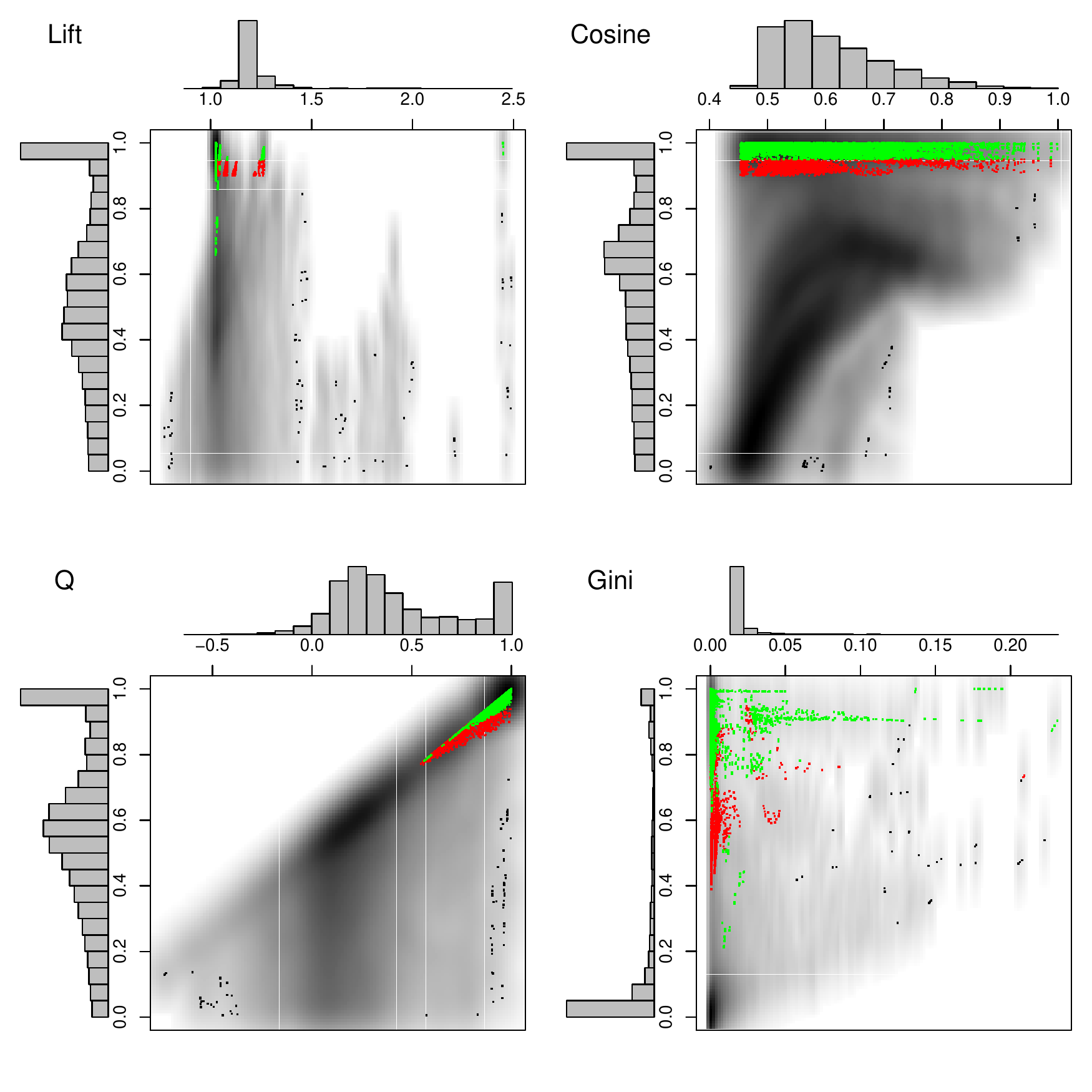}
	\vspace{-0.12in}
\caption{\label{fig:accidents} Plots of interestingness measures against their standardized counterparts for the Belgian accident data. Dark regions indicate a higher density of rules.}
\end{figure}

When comparing the standardized values against raw values, a variety of behaviours are observed (cf.\ Figure~\ref{fig:accidents}). Standardizing $\lift$ results in two very noticeable trends in these rules. The values of $\lift$ and $\slift$ are, generally, positively correlated when the value of $\lift$ is not near one. Rules that have values of $\lift$ near one obtain many values of $\slift$ in the unit interval with a concentration near one.  Even though the cosine similarity is defined in a similar form as lift, no constant value of $\cos$ indicates independence and so the trend observed in the plot with lift is not identifiable. In general, high values of $\cos$ identify positively correlated items and low values of $\cos$ identify negatively correlated items.

The plot of Yule's Q reveals several features. There is a high concentration of values with $\q$ near zero, indicating the same independent rules as for when $\lift$ had a value near one. There are also no rules with low values of $\q$ while achieving high values of $\sq$. This is not particularly surprising because the upperbound of $\q$ is constant. The plot of the Gini index is similar to the others. The majority of rules have values of $\gini$ very near zero. No values of $\gini$ that are high obtain a low value of $\sgini$. In contrast to the other standardized interestingness measures, $\sgini$ maps rules indicating independence to a single value, namely zero, as $\gini$ would already be very close to zero.

In \figref{fig:accidents}, red points highlight rules that obtained values of $\slift$ more than 0.9 and green points highlight rules that obtained values of $\scos$ greater than 0.95. We choose these values to observe how high values of a standardized interestingness measure correspond to other interestingness measures, including the other standardized interestingness measures. In this figure, as well as similar figures in following sections, the raw interestingness measure is on the horizontal axis and the standardized on the vertical axis. In this example, lift and the cosine similarity are specifically selected together because we know that the similar form of $\lift$ and $\cos$ results in similar raw interestingness measures.  Not surprisingly, the relationship between high values of $\slift$ and high values of $\scos$ is strong. These points coincide with relatively high values of $\q$, $\sq$, and even $\sgini$ but not $\gini$. The correspondence of all four standardized interestingness measures speaks to a common type of measurement. This is in contrast to the raw interestingness measures: mostly moderate values of $\lift$, a wide range of values of $\cos$, mostly high values of $q$, and mostly low values of $\gini$. These are reflected more generally in Appendix~\ref{sec:scatterplotmatrices}.

\tabref{tab:accidentsrules} shows rules with the consequent \{35\}. These rules show similar values for an interestingness measure as well as similar values for a standardized interestingness measure. There is, generally, greater disparity in the standardized interestingness measure than in the raw interestingness measure. For example, all of the rules obtained values of $\lift$ of approximately 1.95, but $\slift$ ranged from 0.32 to 0.39 with the lowest values occurring when item \{18\} is in the antecedent. The standardized cosine similarity and standardized Yule's Q exhibit the same behaviour and, to a lesser extent, we observe this with $\sq$. 
\tabref{tab:accidentsRO} contains Kendall's tau-b for these data. Standardizing lift reorders the rankings of rules the most, while standardizing Yule's Q reorders rules the least. The cosine similarity and the Gini index maintain a more moderate relationship when each is standardized. These trends are shown by deciles in Figure~\ref{fig:decilesaccidents} of Appendix~\ref{sec:deciles}. The values of tau-b are fairly close to zero, suggesting a significant reordering of rules for all interestingness measures and all deciles, except for high deciles of Yule's Q that obtain higher values of tau-b. The maintained ordering is most evident in higher deciles of Yule's Q, which may be due to the interestingness measure being the only of the four discussed that always has a constant for an upper bound.
\begin{table}[!h]
\caption{\label{tab:accidentsrules}Rules for the Belgian accident data with the consequent \{35\} sorted by support and confidence.}
\addtolength{\tabcolsep}{-3pt}
\begin{tabular}{c|llllllllll}
\hline
antecedent & supp & conf & raw lift & s. lift & raw cos & s. cos & raw Q & s. Q & raw Gini & s. Gini\\
\hline 
\{12,18,23\} & 0.203 & 0.411 & 1.951 & 0.317 & 0.630 & 0.317 & 0.959 & 0.334 & 0.079 & 0.868\\
\{17,18,23\} & 0.203 & 0.411 & 1.951 & 0.325 & 0.630 & 0.325 & 0.960 & 0.342 & 0.079 & 0.870\\ 
\{18,23\} & 0.203 & 0.411 & 1.951 & 0.327 & 0.630 & 0.327 & 0.960 & 0.343 & 0.079 & 0.870\\ 
\{12,17,23\} & 0.204 & 0.411 & 1.952 & 0.368 & 0.631 & 0.368 & 0.962 & 0.386 & 0.079 & 0.878\\ 
\{12,23\} & 0.204 & 0.411 & 1.952 & 0.370 & 0.631 & 0.37 & 0.962 & 0.388 & 0.079 & 0.878\\ 
\{17,23\} & 0.204 & 0.411 & 1.952 & 0.385 & 0.631 & 0.385 & 0.963 & 0.403 & 0.079 & 0.881\\
\{23\} & 0.204 & 0.411 & 1.952 & 0.387 & 0.631 & 0.387 & 0.963 & 0.405 & 0.079 & 0.881\\
\hline
\end{tabular}\end{table}

\begin{table}
\caption{Rule orders for the Belgian accident data.}
\label{tab:accidentsRO}       
\centering
\begin{tabular}{l|r}
\hline
Interestingness Measure & tau-b  \\
\hline
$\lift$ and $\slift$ &  $-0.033$ \\
$\cos$ and $\scos$ &  $0.380$\\
$\q$ and $\sq$ &  $0.657$\\
$\gini$ and $\sgini$ & $0.250$\\
\hline
\end{tabular}
\end{table}

\section{Reuters}\label{sec:reuters}

The Reuters-21578 Distribution 1.0 data \citep{reuters21578} are well-known data in text categorization. They are a collection of documents appearing on the Reuters newswire in 1987, made available in 1990, and reformatted in 1996. The reformatted version of the data is analyzed here. Words in documents are also reduced to word stems using the \pack{Rstem} package \citep{rstem} for \R~\citep{R10} and stop words are removed. There were over 1,100 key word stems and 17,000 documents containing at least one of these key word stems, with an average of 38 unique key word stems.

\begin{figure}
	\centering\includegraphics[width=0.85\textwidth]{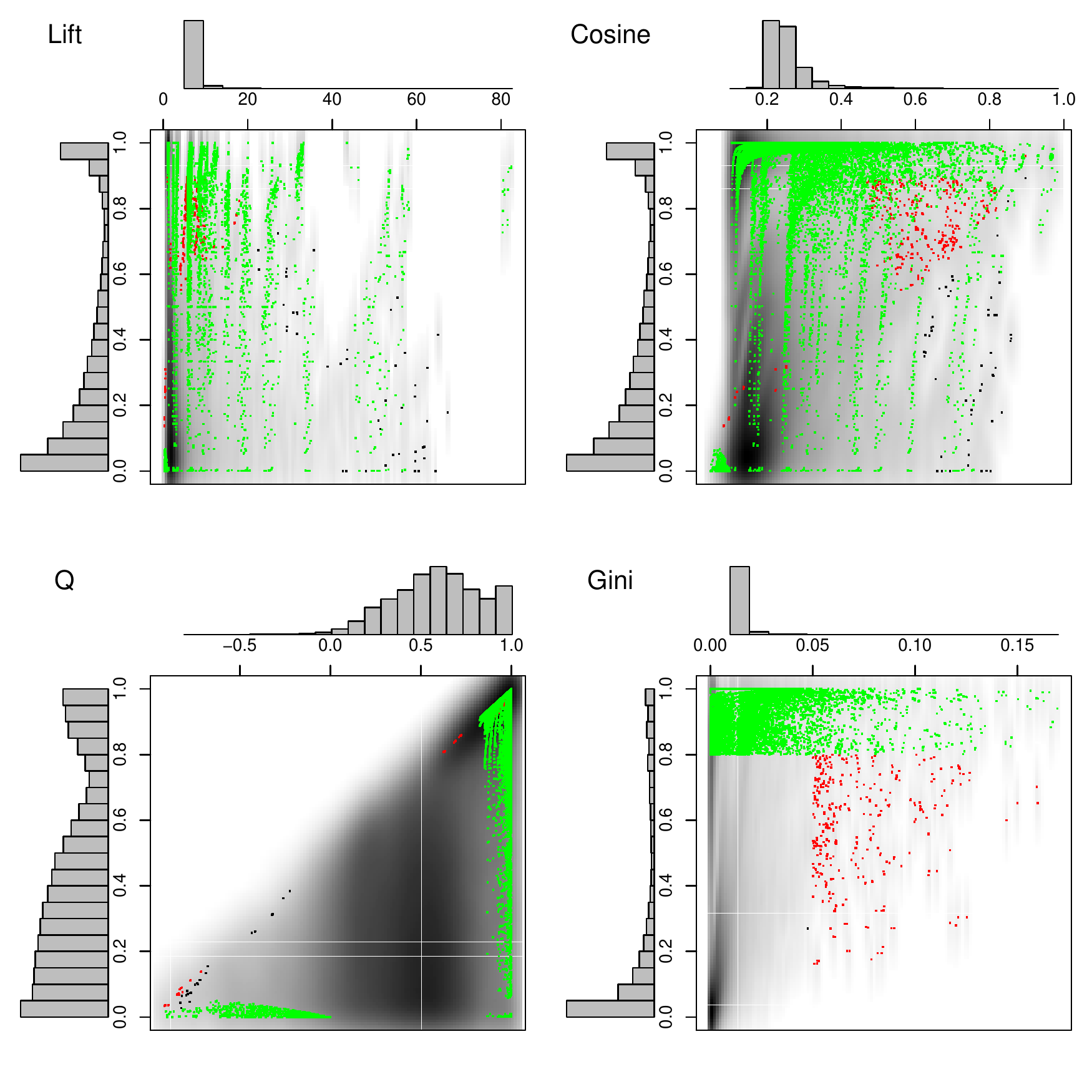}
	\vspace{-0.12in}
\caption{\label{fig:reuters} Plots of interestingness measures against their standardized counterparts for the Reuters data. Dark regions indicate a higher density of rules.}
\end{figure}

Support and confidence thresholds, both set to 0.005, provide over 500,000 rules with the Apriori algorithm. \figref{fig:reuters} shows the plots of raw interestingness measures against their standardized counterparts. The highest values of lift are higher than the previous data, indicating greater dependence. Similar to the previous data, most rules have values of $\lift$ near one and these rules have values of $\slift$ spread throughout the unit interval with concentrations near zero and one. Most rules obtain positive values of $\q$ with these data. When standardized, the range of values of $\sq$ between zero and one is well represented. One concentration occurs at the maximum value of $\q$ near one, which has a large proportion of values of $\sq$ near its maximum of one. The plot of $\sgini$ against $\gini$ shows that most values of $\gini$ are close to zero and correspond to a variety of $\sgini$ values, not just those near zero. However, most of the rules with $\gini$ near zero also have values of $\sgini$ near zero. 


Rules represented in \figref{fig:reuters} are highlighted in red when $\gini$ is above 0.05 and in green when $\sgini$ is above 0.8. High values of $\gini$ correspond to high values of $\cos$ and $\q$, while rules with high values of $\sgini$ correspond to high values of $\lift, \slift, \cos$, and $\scos$ as well as very high values of $\q$ and hence $\sq$. We also observe a few rules that exhibit the opposite behaviour among the negatively correlated rules. This most interestingly manifests in the Yule's Q subplot, which shows the few negatively correlated rules with values of $\q$ below zero. Similar to the previous data, these data indicate high values of $\gini$ related to highly dependent rules. In this second data set, we again observe that values of raw interestingness measures indicating independence have standardized counterparts spread throughout the unit interval rather than being centred around a single value. This shows that while the value of a raw interestingness measure may suggest independence for several rules (Prop.~1 from \secref{sec:interestingness measures}), these rules may have different standardized interestingness measures. 

\tabref{tab:reutersRO} shows values of tau-b for the Reuters-21578 data. Unlike the previous data, lift and the cosine similarity are reordered substantially. Figure \ref{fig:decilesrandom} shows how most rules have been reordered across deciles and interestingness measures.  An example is illustrated in \tabref{tab:reutersrules} for rules containing the consequent \{gold\}. Unlike the previous data, we observe greater differences in raw interestingness measures and even greater differences in the standardized interestingness measures. We also again see that the standardized interestingness measures share some similarities, including generally monotonic relationships with each other (see Figure \ref{fig:reutersmeasures} in Appendix \ref{sec:scatterplotmatrices}). We also observe the near-boundary case for the standardized Gini index. The raw values for the Gini index are all very, very close to zero but the standardized interestingness measures clearly differ from zero in two rules while remaining essentially zero in another rule. Despite being so close to the raw interestingness measure's global lower bound, we see that the standardized interestingness measure for the rule \{reuter\} $\Rightarrow$\{gold\} is relatively high; this shows that this indication of nearly independent rules may not be so close to independence after all. A somewhat similar phenomenon is observed in the previous data in Table \ref{tab:accidentsrules}, where $\sq$ values were moderate even though $q$ values were near the global maximum of one.
\begin{table}[!h]
\caption{\label{tab:reutersrules}Rules for the Reuters Data with the consequent \{gold\} sorted by support and confidence.}
\addtolength{\tabcolsep}{-3pt} 
\begin{tabular}{c|llllllllll}
\hline
antecedent & supp & conf & raw lift & s. lift & raw cos & s. cos & raw Q & s. Q & raw Gini & s. Gini\\
\hline
\{march\} & 0.011 & 0.659 & 1.001 & 0.198 & 0.107 & 0.198 & 0.002 & 0.153 & 3 $\times10^{-8}$  & 7$\times10^{-6}$\\ 
\{said\} & 0.016 & 0.895 & 1.183 & 0.754 & 0.136 & 0.754 & 0.472 & 0.623 & 0.00062 & 0.325\\ 
\{reuter\} & 0.017 & 0.983 & 1.074 & 0.960 & 0.136 & 0.96 & 0.691 & 0.828 & 0.00016 & 0.641\\ 
\hline
\end{tabular}\end{table}
\begin{table}[!h]
\caption{Rule orders for the Reuters data.}
\label{tab:reutersRO}       
\centering
\begin{tabular}{l|r}
\hline
Interestingness Measure & tau-b\\
\hline
$\lift$ and $\slift$ & $-0.264$ \\
$\cos$ and $\scos$ &  $0.023$\\
$\q$ and $\sq$ &  $0.353$\\
$\gini$ and $\sgini$ & $0.234$\\
\hline
\end{tabular}
\end{table}

\section{Random Transactions}\label{random}

The presence of some associations was a reasonable assumption for the two real data sets investigated thus far. We now consider a simulated data set with completely random transactions of independent items. This investigation is necessary to understand how randomness, such as noise, affects standardizing interestingness measures. We expect that many of the raw interestingness measures that fulfil Prop.~1 of \secref{sec:interestingness measures} will achieve values very near the constant, thus implying independence. 
For this reason, we expect no pattern for the values of $\cos$ and, by extension, $\scos$. As in special cases of previous examples, we expect that the standardized counterparts will have no pattern for these rules.

We generate 100,000 random transactions of up to 10,000 items using \pack{random.transactions} from the \pack{arules} package in \R. The probability of items appearing in transactions was left at the default value of 0.01. We then mine the rules with support and confidence thresholds each set to 0.0001  to obtain over 630,000 rules.
\begin{figure}
	\centering\includegraphics[width=0.85\textwidth]{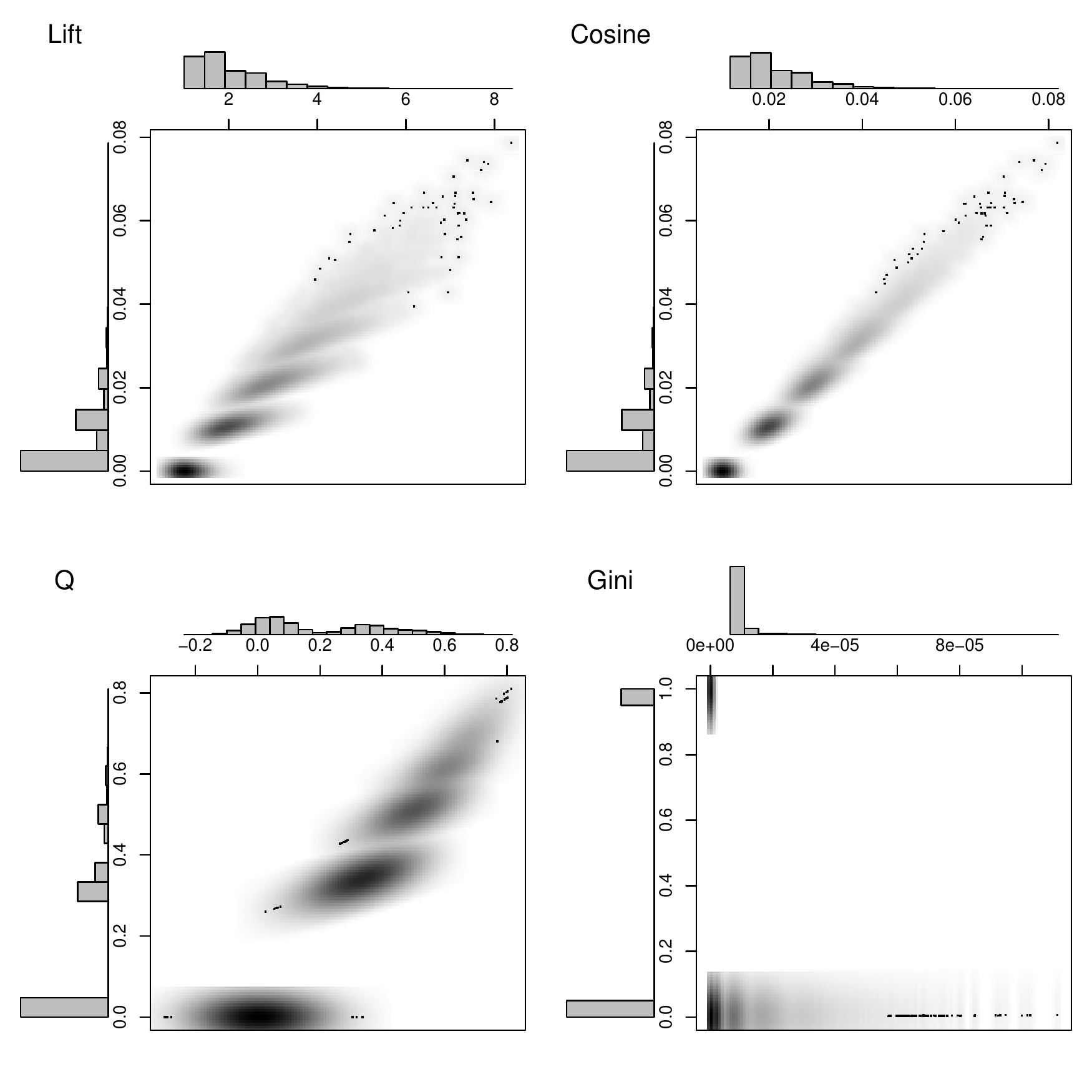}
	\vspace{-0.12in}
\caption{\label{fig:random} Plots of interestingness measures against their standardized counterparts for the random transactions data. Dark regions indicate a higher density of rules}
\end{figure}

\figref{fig:random} shows standardized interestingness measures plotted against their raw counterparts with these random transactions. The scale of each of the plots is important to note: some only show a small range near zero. The `shape' of the plots are similar for the interestingness measures. In three cases, the transformation is nearly linear and the maximum standardized values of $\slift$ and $\scos$ are small. The relationship between $\q$ and $\sq$ is also linear but a larger range is observed, in part because of the constant upperbound, but many of the rules are very close to zero. Contrary to this trend, $\gini$ is very near zero but $\sgini$ polarizes to either zero or one. We investigated this further and the values that tended to one are the entire set of negatively correlated rules. All of the positively correlated rules tended to zero. Though we are observing a dichotomy, whether positively or negatively correlated, we see independent transactions tending to a particular standardized value, depending on the case.

Values of tau-b are shown in \tabref{tab:randomRO}. Though one may have expected that rule orders would appear random with random transactions, this was not the case. The standardized rule orders are surprisingly much more similar to the original interestingness measure rankings compared to previous data. The Gini index was reordered the most, likely because negatively correlated rules score higher than independent rules. Also of interest is the generally monotonically increasing relationship between the decile and value of tau-b for all interestingness measures (see Figure~\ref{fig:decilesrandom} in Appendix~\ref{sec:deciles}).
\begin{table}[!h]
\caption{Rule orders for randomly generated transactions.}
\label{tab:randomRO}       
\centering
\begin{tabular}{c|c}
\hline
Interestingness Measure & tau-b\\
\hline
$\lift$ and $\slift$ &    0.678 \\
$\cos$ and $\scos$ & 0.777\\ 
$\q$ and $\sq$ &  0.809\\
$\gini$ and $\sgini$ &  0.284\\
\hline
\end{tabular}
\end{table}
 
With previous data, we highlighted high values of an interestingness measure. This is not necessary with these data as the transformations appear to be generally monotonic and most rules achieve standardized values very near zero.

\section{Conclusion}

As we have shown through several examples, standardizing interestingness measures serves several purposes not satisfied by raw interestingness measures alone. We have also investigated the behaviour of standardized interestingness measures over different data sets. In the Belgian accidents data, we observed that high values of one standardized interestingness measure correspond to high values of another standardized interestingness measure. Comparing the relative order of rules of a raw against standardized interestingness measure resulted in rules being reordered to various degrees, depending on the interestingness measures; some interestingness measures reordered rules considerably while others maintained much of the original order. 

In the Reuters-21578 data, high values of both the Gini index and standardized Gini index related to relatively high values of the standardized interestingness measures and some of the raw interestingness measures. Exceptions coincide with the natural Gini index scoring high for very negatively correlated data as well as very positively correlated data. The relative order of rules also changed more dramatically with these data, such as when standardizing lift and the cosine similarity. We once again observed greater similarities between standardized interestingness measures than between raw interestingness measures. We also highlighted an example where an interestingness measure took on one of the smallest possible values in its range. The value of the interestingness measure indicated that the antecedent and consequent were very close to being independent, but the standardized interestingness measure was quite high indicating that the value obtained was nearly as far from independence as bounds permit. 

In the last example, we observed how standardized interestingness measures perform on data designed to have no pattern. Although three interestingness measures indicated some form of relationship, standardizing interestingness measures always resulted in values near zero. Only in the case of $\sq$ were values above 0.1 observed. The order of rules by raw interestingness measures was generally similar to the corresponding standardized interestingness measure. Standardizing interestingness measures showed how data with no patterns do not result in an interestingness measure that is misleadingly high whereas the raw interestingness measure alone may have indicated patterns. Standardized interestingness measures often appear to capture the same aspect of a rule by providing similar values; plots in \appref{sec:scatterplotmatrices} show how standardized interestingness measures are more similar to each other than raw interestingness measures. In the cases of $\slift$ and $\sq$, independent rules do not evaluate to a single value but $\sgini$ maps independent rules to zero akin to its raw counterpart; this is not surprising as this is the lower bound of the interestingness measure. Random transactions, on the other hand, showed standardized interestingness measures of rules near zero in most cases.  

One may further study the effects of standardizing interestingness measures on specific properties by considering two families of interestingness measures: one containing interestingness measures possessing the property and the other containing interestingness measures that do not possess the property. A thorough evaluation of several data sets with known patterns and features with members from each of the two families can address more specific aspects of the impact of standardizing interestingness rules. 

\section*{Acknowledgements}
This work was supported by a Discovery Grant from the Natural Sciences and Engineering Research Council of Canada and by an Early Researcher Award from the Ontario Ministry of Research and Innovation (McNicholas).


\appendix
\section{Appendix}

Before deriving any inequalities, recall the following notation:
\newcommand{\minab}{m}
\newcommand{\maxab}{M}
\begin{align*}
	\sigma &~ \text{is the minimum support threshold. When not explicitly given, } \sigma={1}/{n}.\\
	\kappa &~ \text{is the minimum confidence threshold. When not explicitly given, } \kappa={1}/{n}.\\
	\minab &= \min \left\{ \pA,\pB\right \}\\
	\maxab &= \max \left\{ \pA,\pB\right\}
\end{align*}

\subsection{Cosine Similarity}\label{sec:cosderivation}
The lower bounds of $\cos$ are obtained by considering several cases. First, because $\pAB\geq \sigma$,
$$ \cos \geq \frac{\sigma}{\sqrt{\pA\pB}}.$$
Because $\pAB \leq \minab$, $$\cos=\frac{\pAB}{\sqrt{\pA\pB}}=\frac{\pAB}{\sqrt{\minab\maxab}} \leq \frac{\minab}{\sqrt{\minab\maxab}}= \sqrt{\frac{\minab}{\maxab}}.$$ 
We obtain another by considering that $P(A\cup B)=\pA+\pB-\pAB\leq 1$, and so $\pAB\geq\pA+\pB-1$, giving $$ \cos\geq \frac{\pA+\pB-1}{\sqrt{\pA\pB}}.$$
By considering support threshold $\sigma$, if $\pAB\geq \sigma$, the cosine similarity is smallest if $\pA=\pB=(1-\sigma)/2$, giving $$ \cos \geq \frac{2\sigma}{1+\sigma}.$$
If a confidence threshold $\kappa$ is employed, then ${\pAB}/{\pA}\geq \kappa$ and it follows that
$$ \cos \geq  \frac{\pAB}{\sqrt{\frac{\pAB}{\kappa}\pB}} =  \sqrt{\kappa \frac{\pAB}{\pB}},
$$
and because $\pAB\geq \sigma$, we have $$\cos \geq \sqrt{\kappa \frac{\sigma}{\pB}}.$$ 
Also, because $\pAB\geq \kappa\pA$, $$ \cos \geq \frac{\kappa \pA}{\sqrt{\pA\pB}}=\kappa\sqrt{\frac {\pA}{\pB}} .$$
%
Upper bounds of $\cos$ come from considering that $\pAB\leq\minab$ and $\pA\pB=Mm$, giving $$ \cos \leq \sqrt{  \frac{\minab}{\maxab}}.$$

\subsection{Yule's Q}\label{sec:yulederivation}
Yule's Q is given by 
\begin{align}
\notag	\q&=\frac{\pAB P(\overline A, \overline B)-P(A\overline B)P(\overline A B)}{\pAB P(\overline A, \overline B)+P(A\overline B)P(\overline A B)}\\
\notag		&=\frac{\pAB ~[1-\pA-\pB+\pAB] - [\pA-\pAB]~[\pB-\pAB]}{\pAB ~[1-\pA-\pB+\pAB] + [\pA-\pAB]~[\pB-\pAB]}\\
		&=\frac{\pAB-\pA\pB}{\pAB+\pA\pB-2\pAB(\pA+\pB-\pAB)}.
\end{align}
Yule's Q achieves the maximum value of $1$ iff $\pAB=\min\{\pA,\pB\}$ and achieves the minimum value of $-1$ only when $\pAB=0$ or when $P(\overline A B\overline)=0$. It is a strictly monotonically increasing function of $\pAB$ so only these endpoints will be considered.

Given the minimum support threshold $\sigma$, $\pAB\geq \sigma$ so the minimum value of $-1$ is not attained. If, however, $\pAB=\sigma$, then Yule's Q attains the value
\begin{equation}\label{qsupp}
	\frac{\sigma-\pA\pB}{\sigma+\pA\pB-2\sigma(\pA+\pB-\sigma)}.
\end{equation}

Given the minimum confidence threshold $\kappa$, because $\pAB=P(B\mid A) \pA$, we know that $\pAB>\kappa\pA$ and so Yule's Q is greater than the value
\begin{align}
	\notag&\frac{\kappa\pA-\pA\pB}{\kappa\pA+\pA\pB-2\kappa\pA(\pA+\pB-\kappa\pA)}\\
	&= \frac{\kappa-\pB}{\kappa+\pB-2\kappa(\pA+\pB-\kappa\pA)}. \label{qconf}
\end{align}

From \eqref{qsupp} and \eqref{qconf}, the lower bound of Yule's Q is given by

$$
\lambda = \max\left\{ -1, \frac{\sigma-\pA\pB}{\sigma+\pA\pB-2\sigma(\pA+\pB-\sigma)} , \frac{\kappa-\pB}{\kappa+\pB-2\kappa(\pA+\pB-\kappa\pA)}  \right\}.
$$


\subsection{Gini}\label{sec:giniderivation}
Derivation of an alternative form of the Gini index.
	\begin{align}
		\gini\notag&=\pA\left[P(B\mid A)^2+P(\overline B \mid A)^2\right]+\pnotA[P(B\mid \overline A)^2+P(\overline B \mid \overline A)^2]-\pB^2-\pnotB^2 \\
	\notag&=\pA\left[ \frac{\pAB^2}{\pA^2}+\frac{\pAnotB^2}{\pA^2}\right]+\pnotA\left[ \frac{\pnotAB^2}{\pnotA^2}+\frac{\pnotAnotB^2}{\pnotA^2}\right]-\pB^2-(1-\pB^2)\\ 
	\notag&=\left[ \frac{\pAB^2+\pAnotB^2}{\pA}\right]+\left[ \frac{\pnotAB^2+\pnotAnotB^2}{\pnotA}\right]-2\pB^2+2\pB-1\\
	\displaybreak[0]
	\notag&=-2\pB^2+2\pB-2\frac{\pAB[\pA-\pAB]}{\pA}-2\frac{\pnotAB[\pnotA-\pnotAB]}{\pnotA}\\
	\displaybreak[0]
	\notag&=-2\pB^2+2\pB-2(\pB)+2\left[\frac{\pAB^2}{\pA}+\frac{[\pB-\pAB]^2}{1-\pA}\right]\\
	\notag&=\frac{-2\pB^2\pA(1-\pA)}{\pA(1-\pA)}+2\left[\frac{\pAB^2+\pA\pB^2-2\pA\pB\pAB]}{\pA(1-\pA)}\right]\\
	&=\label{usableform} 2\frac{[\pAB-\pA\pB]^2}{\pA(1-\pA)}\\
	&=\frac{2}{1-\pA}[P(B\mid A)-\pB][\pAB-\pA\pB]\label{alternateform}
	\end{align}

Using \eqref{usableform}, we know immediately that the (global) lowerbound of zero is achieved iff  $A$ and $B$ are independent. When $A$ and $B$ are not independent, the Gini index is non-zero.

Using Expression \eqref{usableform}, we can write the Gini index as the product
$$\gini=\frac{2}{\pA(1-\pA)} \times \Big( \pAB-\pA\pB \Big)^2.$$
The first factor is largest when $\pA=\frac{1}{2}$, making it smaller than 8. The second term can be no larger than $\frac{1}{16}$, precisely when $A$ and $B$ are most positively correlated. Let \mbox{$z=\pA=\pB=\pAB$} in this instance,
\begin{align*}
	(\pAB-\pA\pB)^2\leq (z-z^2)^2 \leq \left(\frac{1}{2}-\left(\frac{1}{2}\right)^2\right)^2=\frac{1}{16}.
\end{align*}
From these deductions, we know that
\begin{equation}\label{globalupper}
	\gini \leq 8\times \frac{1}{16}=\frac{1}{2}.
\end{equation}

\newcommand{\lowestAB}{l}

We may obtain other inequalities by considering the highest and lowest values of $\pAB$ influences the numerator of \eqref{usableform}. Let $\lowestAB=\max \{ \sigma, \pA+\pB-1, \kappa\pA\}$. From \secref{sec:cosderivation}, $l\leq\pAB$ and $\pAB\leq \minab$, as before. 

Due to the quadratic nature of the numerator, we must be slightly more cautious. When $\pAB-\pA\pB\geq0$, 
\begin{equation}\label{giniupper1}
	2\frac{(\lowestAB-\pA\pB)^2}{\pA(1-\pA)}\leq \gini. 
\end{equation}

When $\pAB-\pA\pB>0$, the above inequalities are reversed to obtain \begin{equation}\label{giniupper2}2\frac{(\minab-\pA\pB)^2}{\pA(1-\pA)}\leq \gini. 
\end{equation}

When $\pAB<\pA\pB$, we can also employ $\pAB\geq \kappa \pA$ to obtain
$$\gini\geq 2\frac{ (\pAB-\pA\pB) (\kappa-\pB)} {1-\pA},
$$
and when $\pAB>\pA\pB$, we find that
$$\gini\geq 2\frac{(\pAB-\pA\pB)(1-\pB)}{1-\pA}.
$$

\pagebreak
\subsection{Scatterplot Matrices}
\label{sec:scatterplotmatrices}
\begin{figure}[!ht]
  \subfigure[Raw Interestingness Measures]{%
  	\includegraphics[width=.5\textwidth]{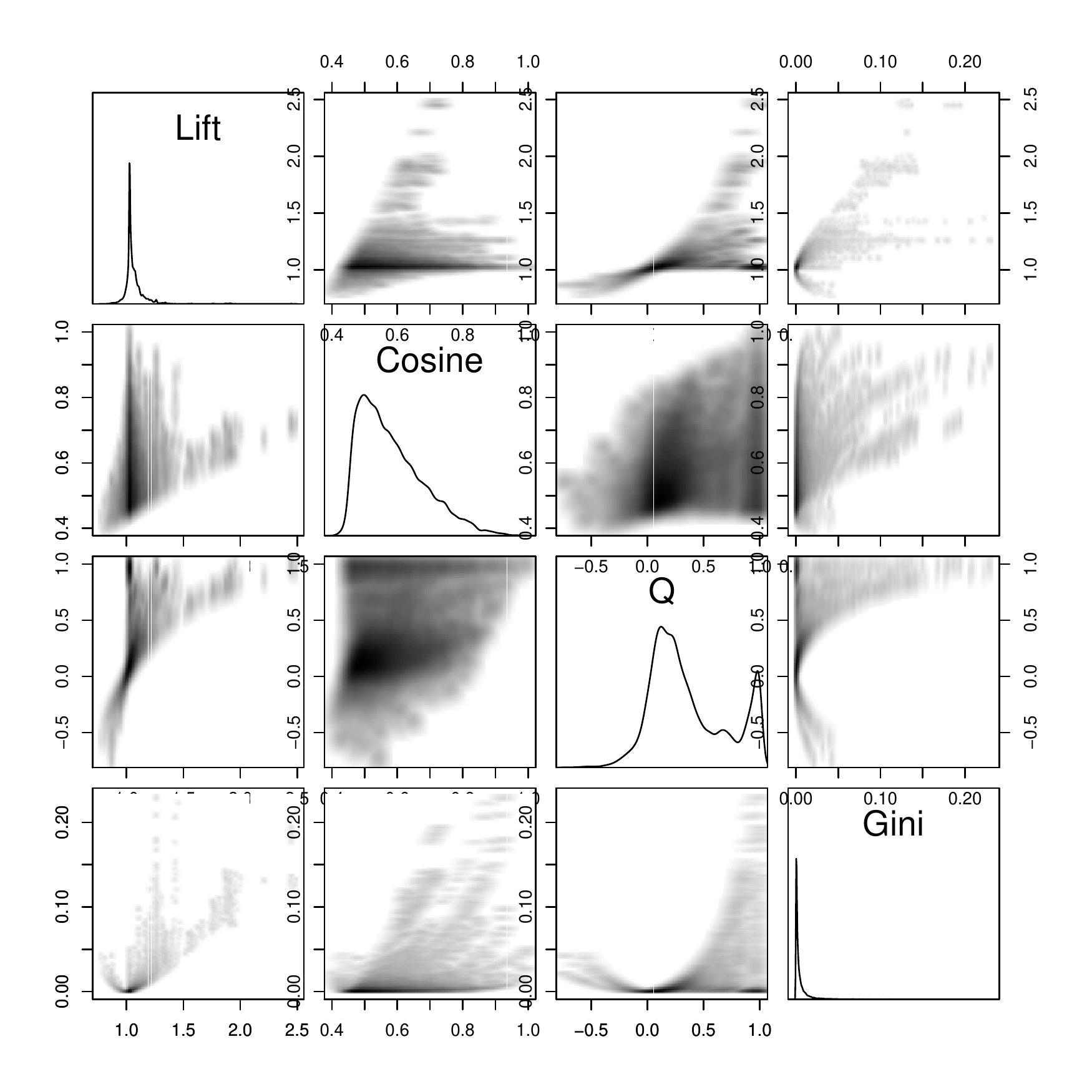}
  }  \subfigure[Standardized Interestingness Measures]{%
	\includegraphics[width=.5\textwidth]{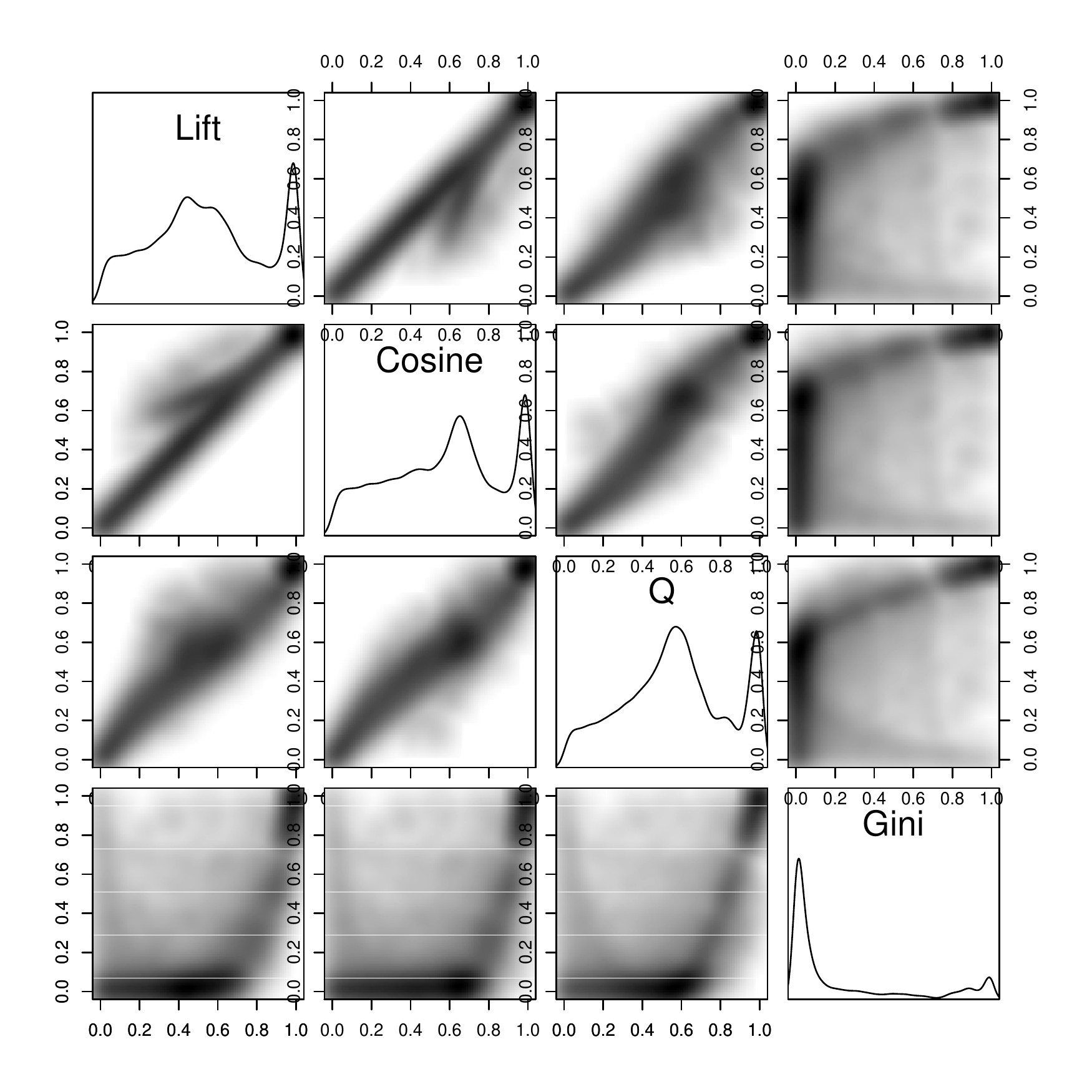}
  }
\caption{\label{fig:accidentsmeasures} Scatterplot matrices of interestingness measures for the Belgian accident data.}
\end{figure}

\begin{figure}[!ht]
  \subfigure[Raw Interestingness Measures]{%
  	\includegraphics[width=.5\textwidth]{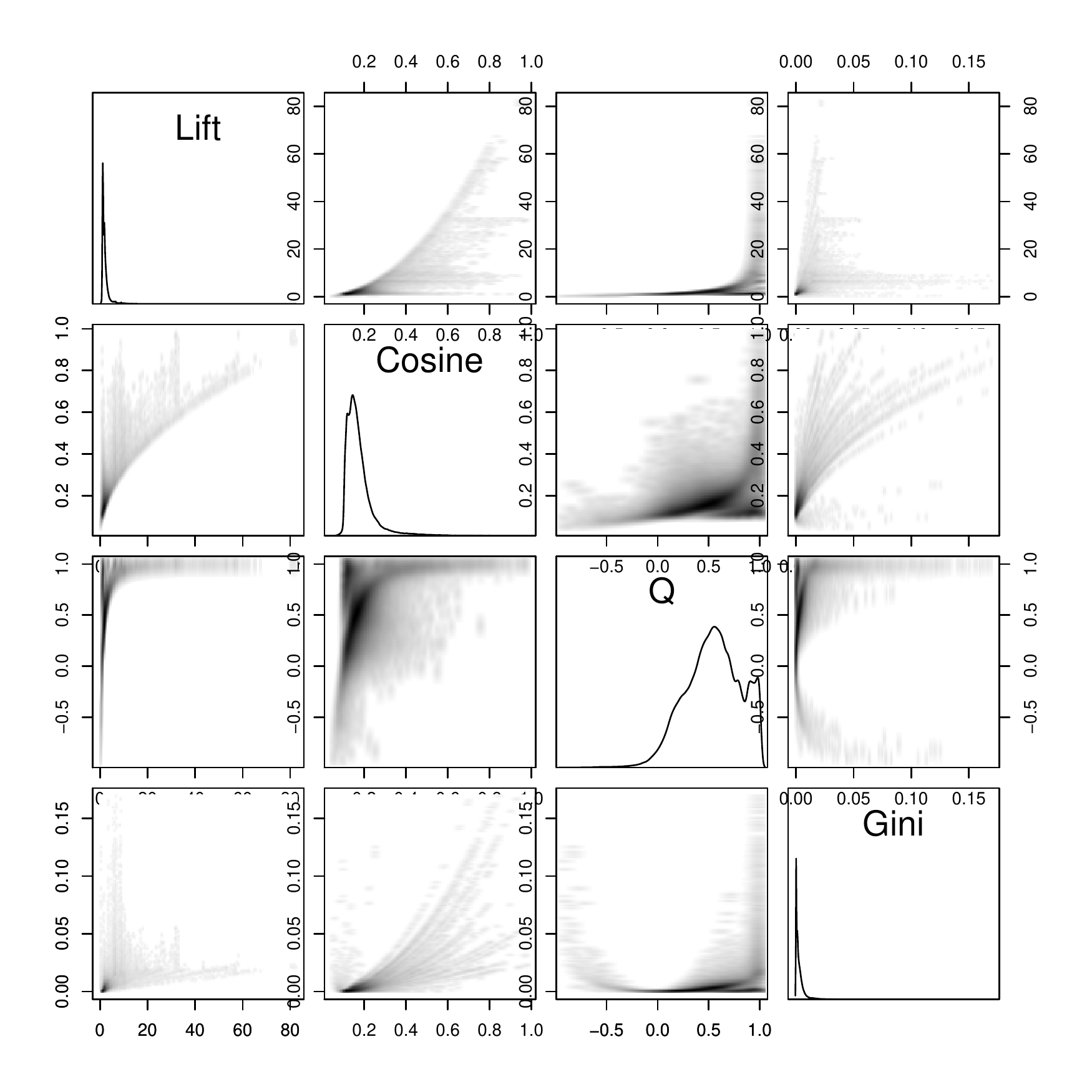}
  }  \subfigure[Standardized Interestingness Measures]{%
	\includegraphics[width=.5\textwidth]{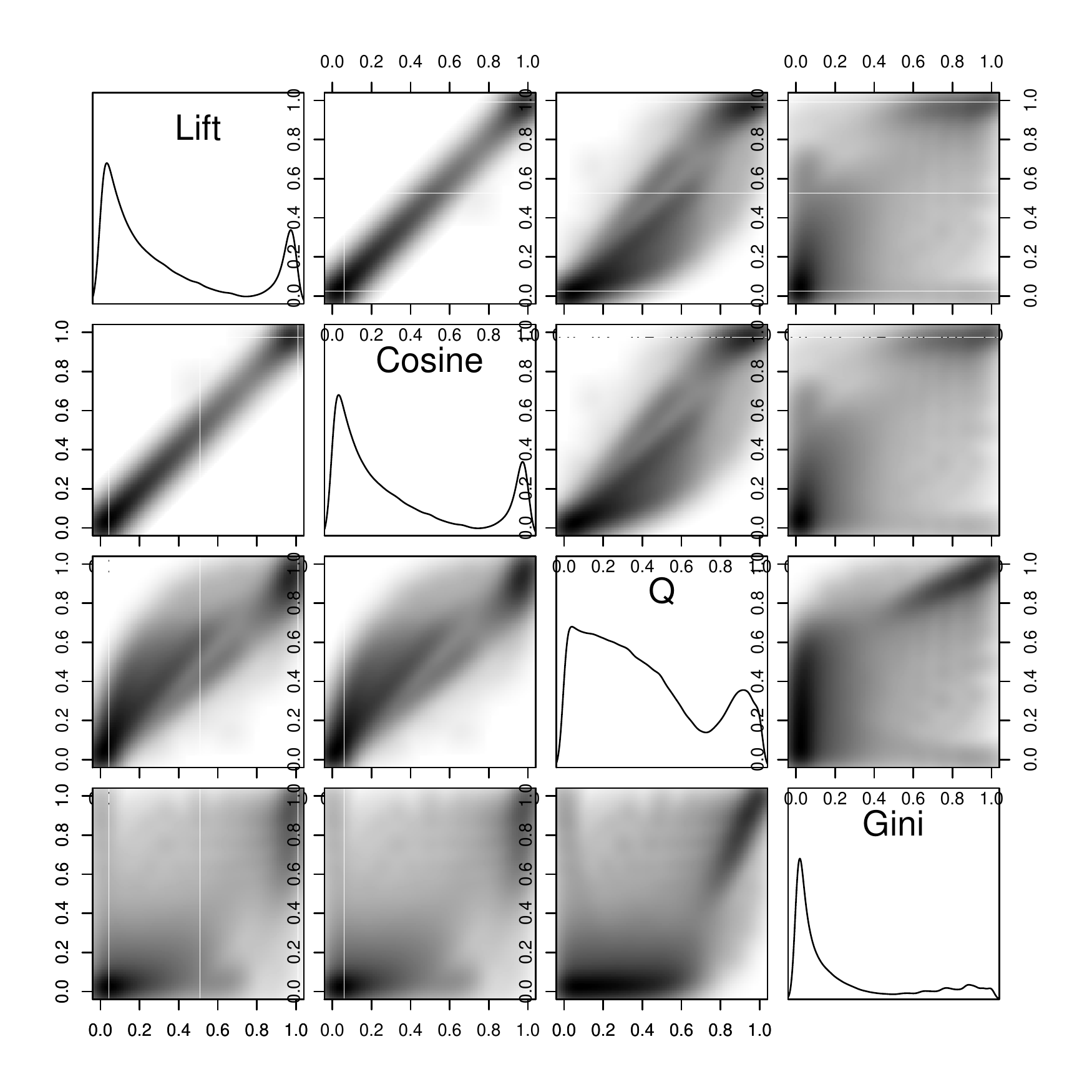}
  }
\caption{\label{fig:reutersmeasures} Scatterplot matrices of interestingness measures for the Reuters data.}
\end{figure}

\begin{figure}[!ht]
  \subfigure[Raw Interestingness Measures]{%
  	\includegraphics[width=.5\textwidth]{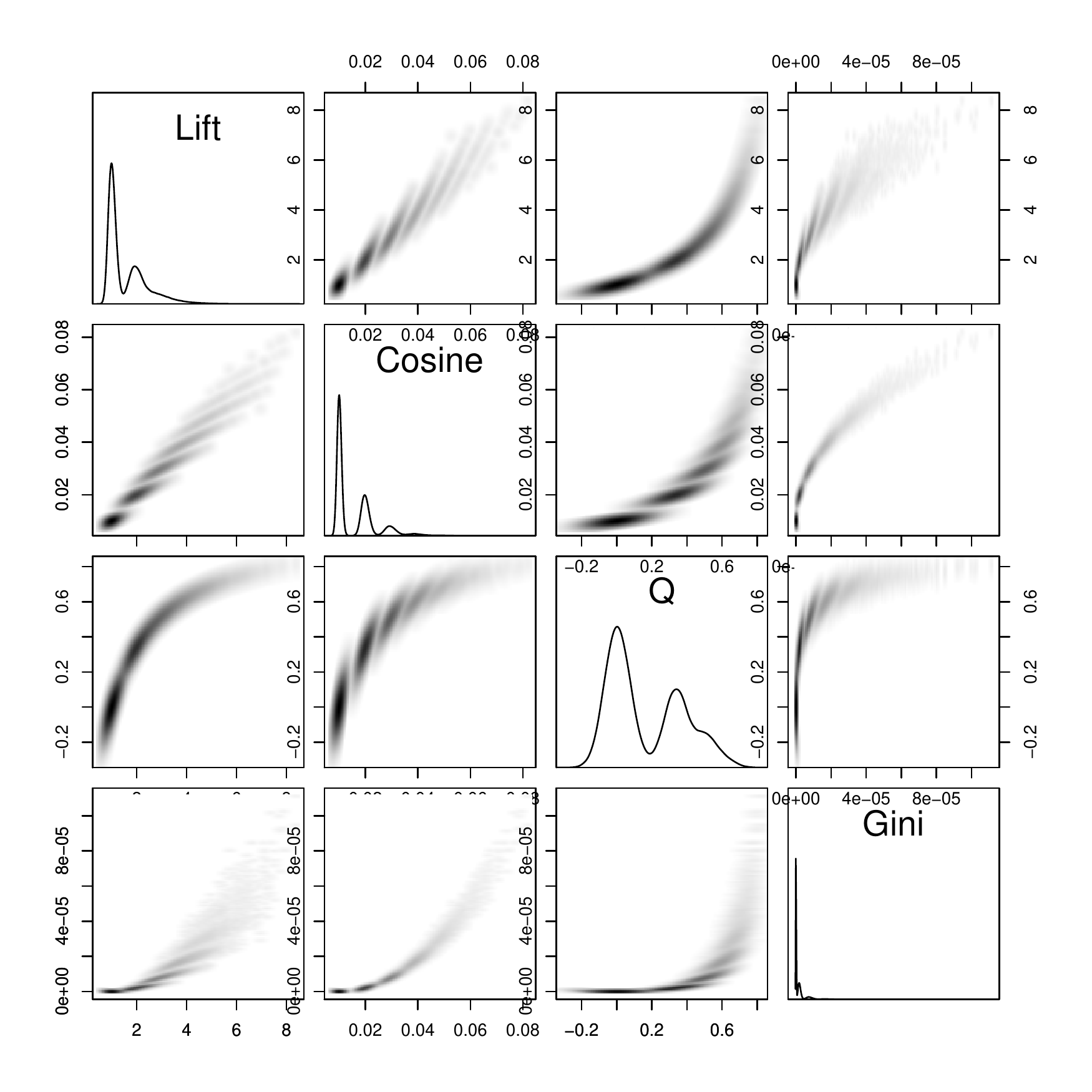}
  }  \subfigure[Standardized Interestingness Measures]{%
	\includegraphics[width=.5\textwidth]{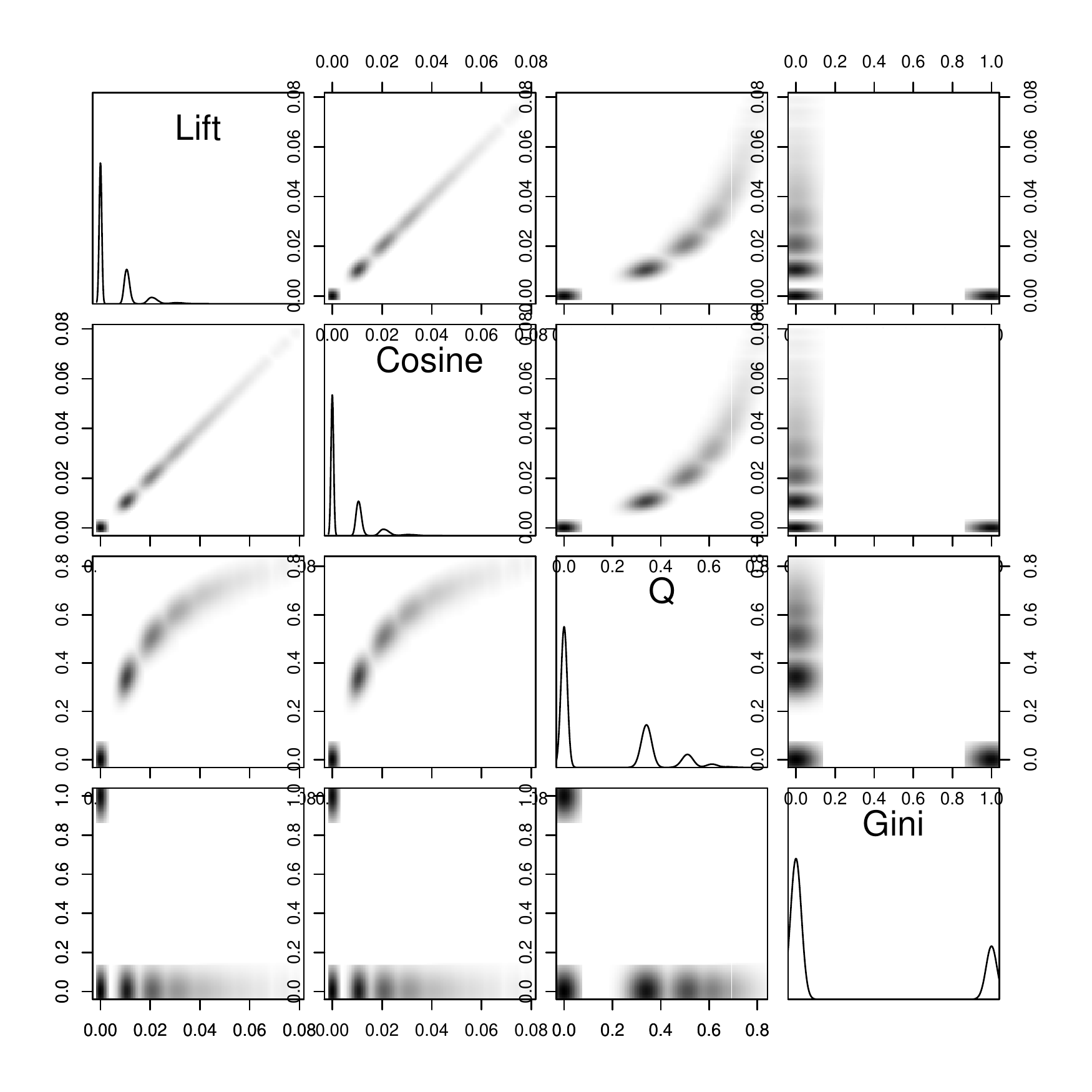}
  }
\caption{\label{fig:randommeasures} Scatterplot matrices of interestingness measures for random transactions.}
\end{figure}

\vspace{2em}

\newpage
\subsection{Tau-b by Deciles}\label{sec:deciles}
\begin{figure}[!ht]
\centering
	\subfigure[Accidents Data]{
  		\includegraphics[width=0.6\textwidth]{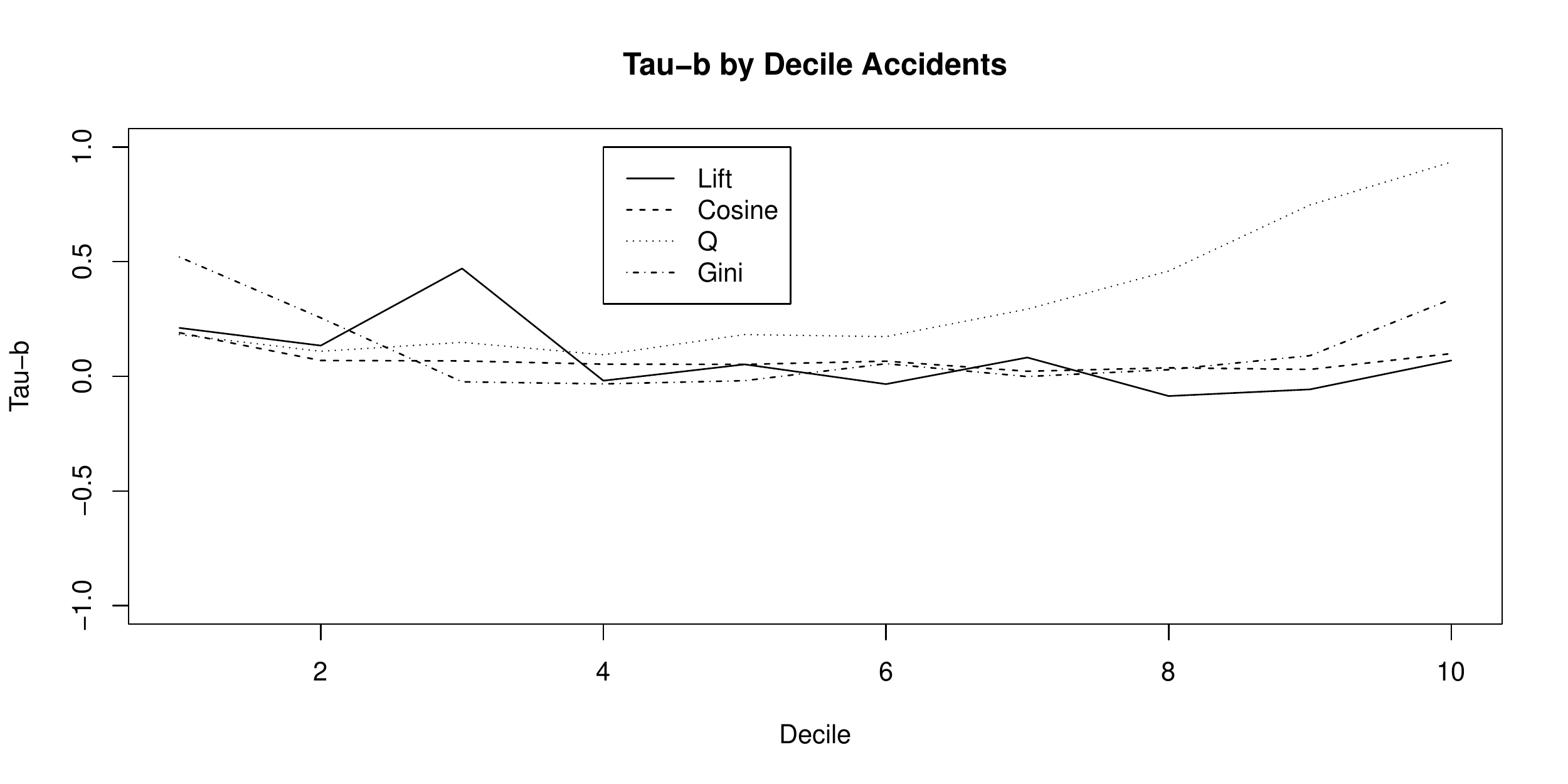}
		\label{fig:decilesaccidents}
	}\\
	\subfigure[Reuters-21578 Data]{
  		\includegraphics[width=0.6\textwidth]{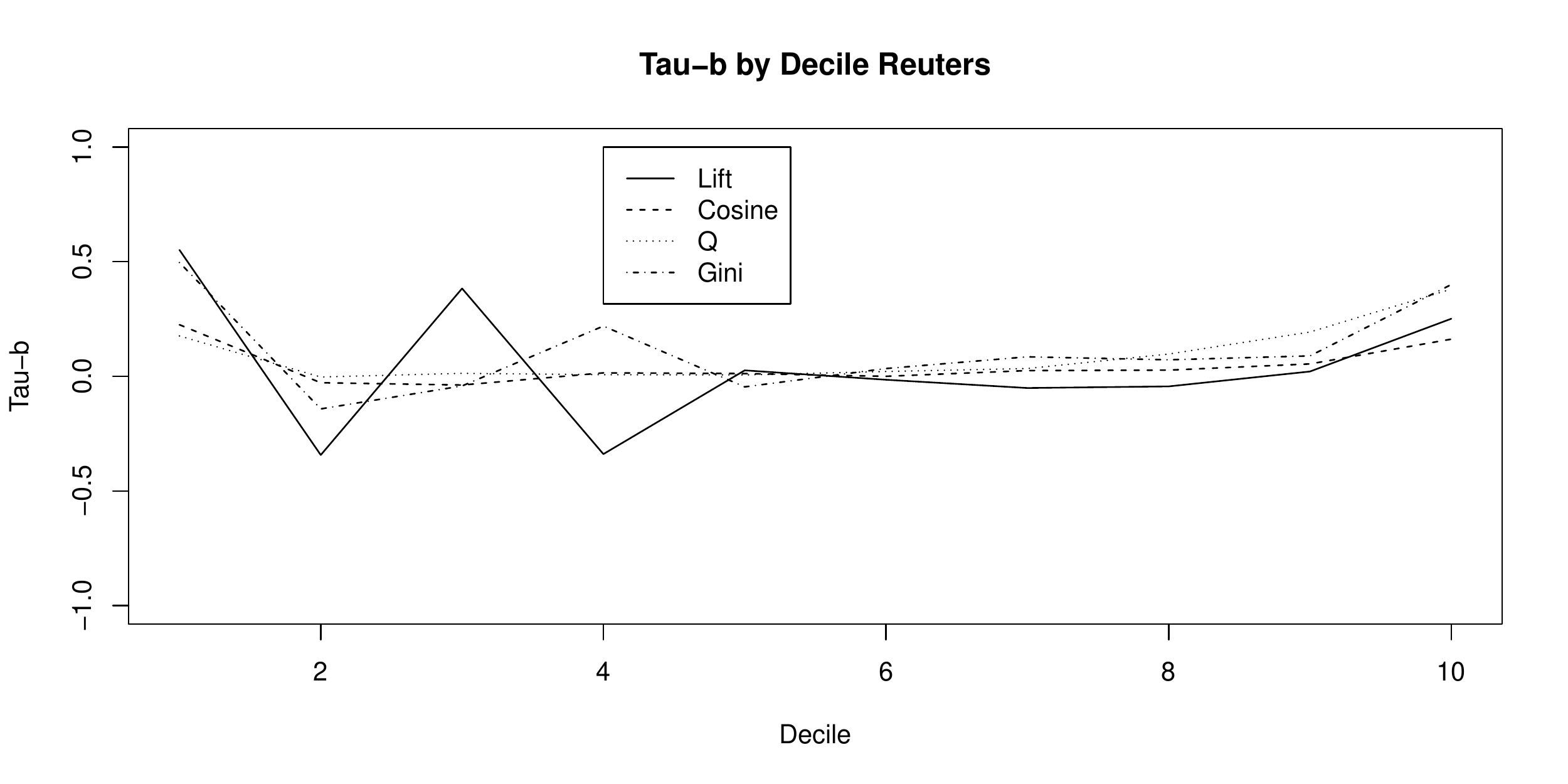}
		\label{fig:decilesreuters}
	}\\
	\subfigure[Random Data]{
  		\includegraphics[width=0.6\textwidth]{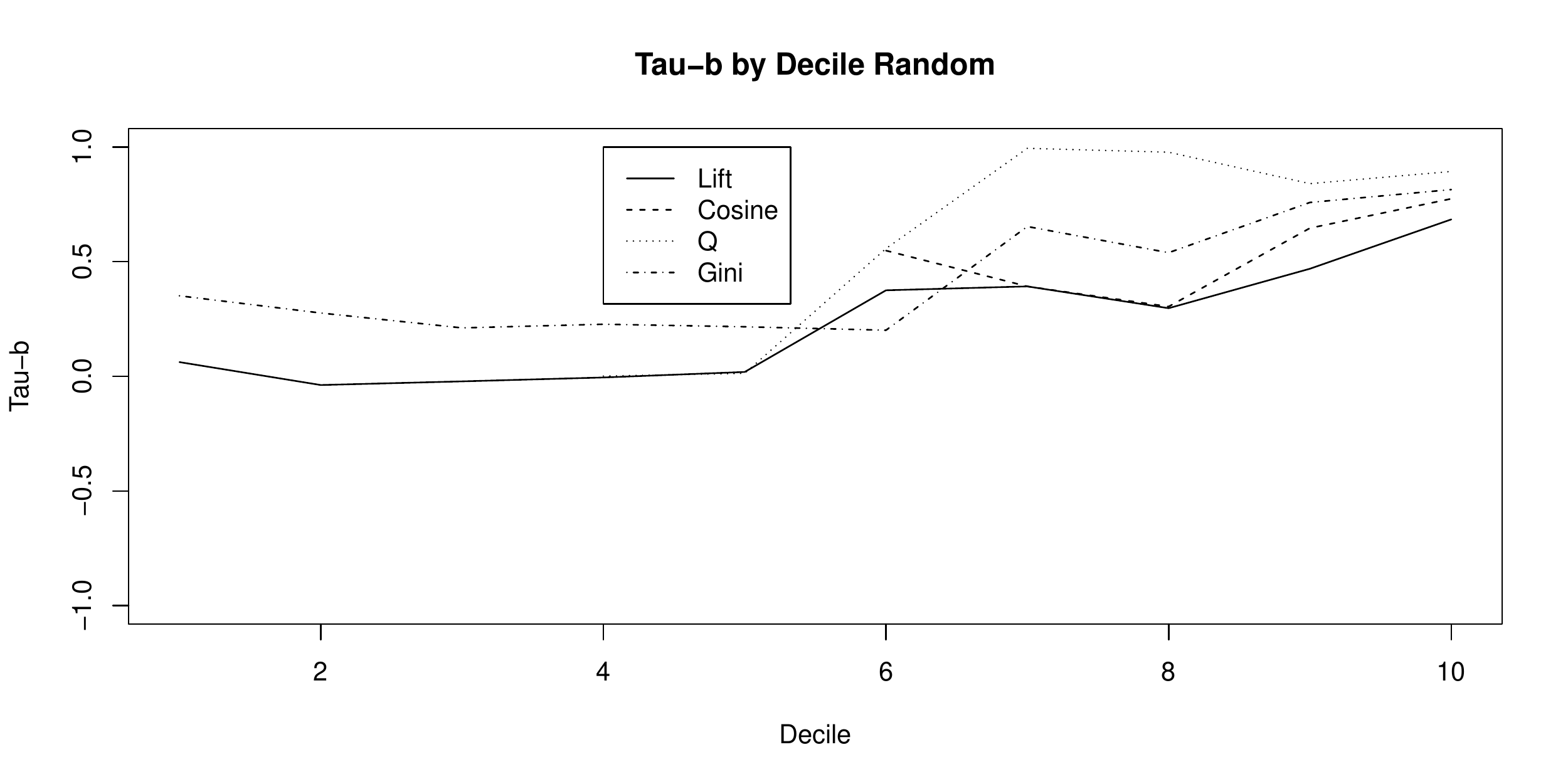}
		\label{fig:decilesrandom}
	}
\caption{\label{fig:deciles} Tau-b by deciles for each data set.}
\end{figure}

\end{document}